# Dynamics of Polymer Chains in Disperse Melts: Insights from Coarse-Grained Molecular Dynamics Simulations

Taofeek Tejuosho[1], Sohil Kollipara[2], Sumant Patankar[1], Janani Sampath[1]

[1]Department of Chemical Engineering, University of Florida, Gainesville, Florida 32611

[2]Department of Chemistry, University of Florida, Gainesville, Florida 32611

*jsampath@ufl.edu

**Abstract**

Synthetic polymers have a distribution of chain lengths which can be characterized by dispersity, Đ. Their macroscopic properties are influenced by chain mobility in the melt and controlling Đ can significantly impact these properties. In this work, we present a detailed study of the static and dynamic behavior of fully flexible polymer chains that follow Schulz – Zimm molecular weight distribution up to Đ = 2.0 using coarse-grained molecular dynamics simulations. We analyze the behavior of test chains with molecular weights that are equal to, above, or below the molecular weight ($M_w$) of the melt. Static analysis shows that the conformation of these test chains remains unaffected by the heterogeneity of the surrounding chains. To study the dynamics, we compute the mean squared displacement of test chains in melts of the same $M_w$ and different dispersities. The mobility of test chains with $N > M_w$ steadily increases with dispersity, due to the shorter chains contributing to early onset of disentanglement of the long chains. However, the dynamics of test chains of length $N < M_w$ is non – monotonic with respect to dispersity, this behavior arises from a tradeoff between the increased mobility of shorter chains and the corresponding slowdown caused by the presence of longer chains. We examine the dynamic structure factor and find a weakening of tube confinement, with the effects becoming less pronounced with increasing dispersity and $M_w$. These findings provide insights into the rich dynamic heterogeneity of disperse polymer melts.

**Introduction**

Manipulating polymer properties without altering monomer chemistry or by using external additives is a desirable strategy to improve their processability and mechanical performance, and is an active area of research in polymer science.[1,2] A central characteristic of synthetic polymers is the molecular weight distribution (MWD) or the distribution of chain lengths in a given sample, which is a consequence of the statistical nature of the polymer synthesis routes. Dispersity (Đ) is a measure of the extent of this distribution and mathematically, it is the ratio of weight average ($M_w$) and number average ($M_n$) molecular weight.[3] Controlling Đ is a promising technique to tune polymer properties without changing monomer composition and plays a central role in influencing rheology, mechanical properties, and phase behavior of melts, which in turn dictates their final application use.[2,4–9] While much of the focus in the past has been to obtain a narrow MWD polymers with low Đ, it has been recently shown that polymers with moderate (Đ = 1.2–1.5) and high dispersity (Đ > 1.5) are desirable as they provide a balance of processability and mechanical strength.[9–13] This has led to an increase in the development of various synthetic methods to precisely target different Đ values, both high and low.[5,8,9,14,15]

Most commodity polymers, due to their high molecular weights and Đ, contain both long chains that are well-entangled and short chains that are below the entanglement limit. These chain entanglements – topological constraints imposed by surrounding chains on a test chain, restricting its motion, dictate the dynamics of polymer melts.[16,17] Theoretically, dynamics of short unentangled chains in a melt are well described by the Rouse model.[17,18] However, when the chains are sufficiently long, entanglements become significant. The tube model by Edwards and de Gennes, refined by Doi and Edwards through the reptation theory, has found considerable success in describing dynamics of entangled chains in a monodisperse melt at the intermediate time and length scales.[19–22] Studies have shown that dynamics of linear chains in disperse melts cannot be described solely by the classical reptation theory due to additional relaxation processes such as contour length fluctuation and constraint release.[21,23,24] Double reptation theory by Cloizeaux captures the heterogeneity in chain sizes by assuming that the relaxation dynamics of each chain species in the melt is unaffected by the presence of other chains.[25] However, it has since been shown that the dynamic behavior of chains is dependent on composition of the chain in the melt.[26,27,28] Zamponi et. al showed the effect of constraint release mechanism on chain relaxation of binary polymer blends using neutron spin echo (NSE) experiments.[29] They found that decreasing the chain length within the matrix reduces entanglements and long chains exhibit unconfined Rouse motion. As polydispersity is inevitable for synthetic polymers, studying the interplay due to variability in chain length within the melt and its influence on their relaxation dynamics, which in turns drives observable properties such as mechanical and viscoelastic responses may have both fundamental and industrial implications.

Coarse-grained molecular dynamics (MD) simulations have been used extensively to elucidate the structure and dynamics of entangled polymer melts.[4,28,30–36] Kröger and Voigt showed that the dynamical crossover from Rouse to Reptation diffusion regime scales as $D \sim N^{-1}$ to $D \sim N^{-2}$, respectively, consistent with the findings of Kremer and Grest.[32] Despite fundamental and industrial significance of disperse melts, there have only been a handful of studies that have used computational methods to study their structure and dynamics.[4,37–40] Baschnagel et. al,[38] used the bond fluctuation model for bidisperse polymer melts to probe the structure and dynamics in both athermal and thermal conditions. Oluseye et al.,[39] investigated the effects of variability of chain lengths on dynamics of simple bidisperse melts, and found that the mobility of short chains qualitatively remains the same in the presence of long chains, however, their presence significantly accelerate the dynamics of the long chains. Dorgan et al. provided a methodology for mapping various MWD using a lattice-based Monte Carlo approach in which the chains were represented by connected sequences of lattice segments. They showed that for the same $M_w$, disperse polymers have a lower Rouse relaxation time with a broadening of the reptation transition, compared to monodisperse systems.[41] Recently, Peters et al. probed the effects of dispersity on chain mobility for relatively narrow MWD (low Đ) melts using a chemically specific coarse-grained model.[4,37] They found that although Đ in the range of 1.0 - 1.16 does not affect entanglement time or tube diameter, the shortest chains in a disperse melt move significantly faster than the longest ones, thereby offering a constraint release pathway for the longer chains.

In this paper, we investigate how dispersity influences dynamics in weakly to moderately entangled polymer melts by using the Kremer – Grest (KG) bead-spring model[34] and reveal the often missing physics due to variability in chain lengths. We account for Đ by generating initial configurations that follow Schulz – Zimm MWD,[42–44] with Đ = 1.0 – 2.0 and $M_w$ = 60 – 360, which are considerably higher than the Đ and $M_w$ used in previous studies. This ensures that we capture the behavior of chains with distinct dynamic regimes, which may not be observable in melts of small $M_w$ and narrow chain length distributions (low Đ). We describe static conformational structures of the polymer chains and compare them with those of ideal chains. Next, we conduct detailed analysis of the dynamic behavior of test chains in the melt by characterizing the mean-squared displacement, coherent dynamic structure factor, and polymer end-to-end vector autocorrelation function. The structure factor analysis in particular allows us to capture the molecular mechanisms that occur due to this rich heterogeneity in chain lengths, significantly different from prior work. We also provide insights into entanglement dynamics in melts with high $M_w$ and Đ, which is different from the low $M_w$ and narrow Đ systems considered previously.

**Methods**

We model fully flexible chains in a melt using the KG bead – spring model.[34] Chains are represented as a sequence of beads connected by finitely extensible, non – linear elastic (FENE) springs. The potential between the bonded beads is:

$$U_{FENE}(r) = -\frac{1}{2}KR_0^2 \ln\left[1 - \left(\frac{r}{R_0}\right)^2\right] + 4\epsilon\left[\left(\frac{\sigma}{r}\right)^{12} - \left(\frac{\sigma}{r}\right)^6\right] + \epsilon, \quad (1)$$

where r is the distance between the beads, $\sigma$ and $\epsilon$ are the standard Lennard Jonnes (LJ) units of length and energy, respectively. The LJ unit of time is $\tau = (m\sigma^2/\epsilon)^{1/2}$, where m is the monomer mass. The first term of equation 1 captures the attractive potential between bonded monomer beads, which extends to $R_0$, the maximum extent of the bond with magnitude $1.5\sigma$, the standard value. The second term of equation 1 accounts for the excluded volume repulsion between beads. Standard value of $K = 30\sigma/\epsilon$ is chosen to prevent chain crossing. The interaction between non – bonded monomers is modelled with the Lennard Jones potential given as:

$$U_{LJ}(r) = 4\epsilon\left[\left(\frac{\sigma}{r}\right)^{12} - \left(\frac{\sigma}{r}\right)^6\right] \quad (2)$$

This potential is truncated at $r = 2.5\sigma$ to account for repulsion at short distance and attraction at longer distance, and it is shifted to zero beyond the cutoff.[28] We use GPU-accelerated large-scale atomic/molecular massively parallel simulator (LAMMPS) MD simulation package[45] with a velocity Verlet algorithm and a time step $t = 0.005\tau$, as used in prior CG studies.[46,47] Initial configurations were generated by randomly placing a specified number of chains in a simulation box at a density $\rho = 0.85$ and mean bond length $l_b = 0.97$.[48] As the chains are placed through a constrained random walk in the simulation box, any overlapping beads need to be pushed off before the pairwise potentials are turned on so that the system does not blow up. To do this, we employ a soft potential[45,48] given by equation S1 in the Supplementary Information.

The temperature used in this study is $T = 1.0\epsilon/k_B$, ($k_B$ is the Boltzmann constant) and was kept constant by weakly coupling the polymer beads to a heat bath with a Nosé-Hoover thermostat, with a temperature damping parameter of $T = 1\tau$. A barostat with a damping parameter of $100\tau$ was used to maintain the system pressure of 0. The LJ potential was turned on during equilibration at constant temperature and pressure for a simulation time which depends on dispersity and total number of beads, $N_t$. Production runs up to $10^9$ timesteps is used in this study for the largest melt, $M_w = 360$. Equilibration is confirmed through decorrelation of the average end-to-end vector autocorrelation function $C_{ee}(t)$ of the bulk and the longest 20% chains in each system to a value < 0.1.

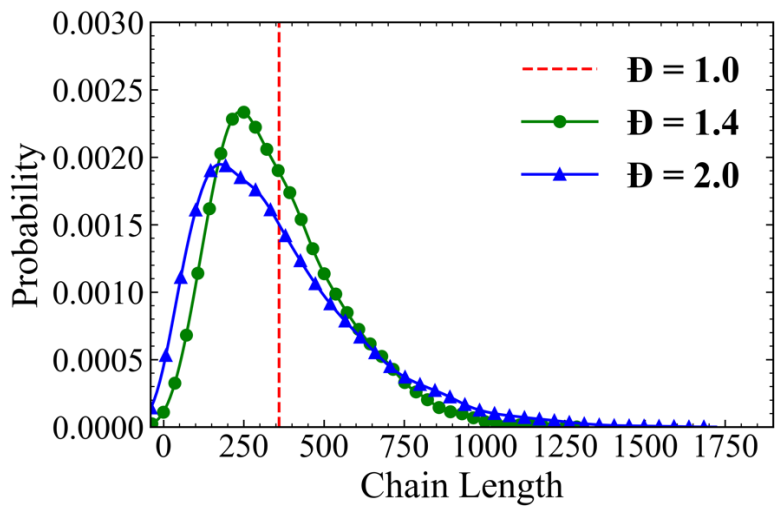

**Figure. 1. Distribution of chain lengths that follow Schulz – Zimm molecular weight distribution for $M_w = 360$ with Đ = 1.0 (red dashed line), 1.4 (green) and 2.0 (blue).**

To capture the effects of dispersity, we use the Schulz – Zimm distribution, which represents the distribution of chains synthesized by chain – growth polymerization, such as in anionic and atom-transfer radical polymerization.[42–44,49] It is mathematically represented by:

$$P(M) = \frac{z^{z+1}}{\Gamma(z+1)} \frac{M^{z-1}}{M_n^z} \exp\left(\frac{-zM}{M_n}\right) \quad (3)$$

Where $\Gamma(z)$ is the gamma function, z is the Schulz – Zimm parameter which determines the polydispersity index, $Đ = \frac{M_w}{M_n} = \frac{z+1}{z}$, $M_n$ is number average molecular weight and $M_w$ is weight average molecular weight. For illustration purposes, Figure 1 shows the distribution for a melt of $M_w$ = 360, Đ = 1.0, 1.4 and 2.0. The distribution for the other $M_w$ considered can be found in Figure S1 of the Supplementary Information. By only varying the parameter z, it is possible to obtain reasonable descriptions for narrow and moderately broad MWD.[43] As shown in Table 1, our design space spans unentangled to moderately entangled chains. For our largest system (Đ = 2.0 and $M_w$ = 360), we have 4000 total chains to accurately capture chain length statistics with over 650 unique chains to capture the non – uniformity in chain lengths. In comparison with prior studies that have used KG model for probing dispersity effects on polymer melts properties, the chains in the current study are significantly longer. For example, in the melt of $M_w$ = 360 and Đ = 1.4, we have more than 450 chains (from a total of 2000 chains) that are greater than N = 360 (N is the chain length which corresponds to the number of monomeric units or "beads") in the melt. We independently generate three replicas for each dispersity to ensure better statistics, and accurate sampling of chain lengths in the highly disperse melts. The three independent runs produce consistent results, and the simulation data presented in this work are averaged over the three runs.

**Table 1.** Dispersity Đ, weight average molecular weight $M_w$, total number of chains in melt $N_c$ and total number of beads (monomers) $N_t$

| Đ | $M_w$ | $N_c$ | $N_t$ |
|---|---|---|---|
| 1.0 | 60 | 1000 | 60,000 |
|  | 140 |  | 140,000 |
|  | 220 |  | 220,000 |
|  | 360 |  | 360,000 |
| 1.4 | 60 | 2000 | 86,499 |
|  | 140 |  | 203,082 |
|  | 220 |  | 318,244 |
|  | 360 |  | 506,635 |
| 2.0 | 60 | 4000 | 121,244 |
|  | 140 |  | 276,554 |
|  | 220 |  | 436,329 |
|  | 360 |  | 726,408 |

**Results**

We first study global conformational properties of test chains by estimating their mean square end-to-end distance $\langle R_{ee}^2 \rangle$ and the radius of gyration $\langle R_g^2 \rangle$. We calculate the conformation of chains that are within ±5 of N, where N is the chain length that corresponds to the weight average molecular weight of the melt $M_w$. Results of $\langle R_{ee}^2 \rangle$ and $\langle R_g^2 \rangle$ vs. N are shown in Figure 2a. According to Flory's theory, for ideal chains in a monodisperse melt, $\langle R_{ee}^2 \rangle / \langle R_g^2 \rangle \approx 6$ and $\langle R_{ee}^2 \rangle \propto \langle R_g^2 \rangle \propto N^{2\nu}$, where $\nu \approx 0.5$ is the Flory exponent.[50–52] We observe that the test chains maintain the same average size in both the disperse and uniform melts. This indicates that the surrounding changes caused by neighboring chains of different lengths do not affect the global conformation of the test chain. These findings are consistent with previous studies of Baschnagel et. al.[38], as the $R_{ee}$ and $R_g$ of chains in monodisperse and bidisperse melts coincide. We also characterize the conformational statistics through the normalized probability distribution of the end-to-end distance $h_N(r_e)$, where $r_e = (R_{ee}^2/\langle R_{ee}^2 \rangle)^{1/2}$. $\langle R_{ee}^2 \rangle$ is the average squared end – to – end distance of test chain of size N. We fit our simulation data to the normalized theoretical prediction of $R_{ee}$ for ideal chains,[36] which is a Gaussian distribution, given by

$$G_e(r_e) = 4\pi r_e^2 \left(\frac{3}{2\pi}\right)^{3/2} exp\left(-\frac{3r_e^2}{2}\right). \qquad (4)$$

We find that the distribution of chains in the monodisperse melt is Gaussian, in good agreement with the theoretical prediction, as shown in Figure 2b. For disperse systems, we perform this analysis for test chains and find their distribution to also be Gaussian. When we repeat this analysis for all chains in a melt, we find that the distribution of chains in the disperse melts are skewed to the right with a long tail which increases with Đ. Large number of short chains skew the distribution to the right, while the long chains give rise to the tail, both characteristics reflective of the Schulz – Zimm distribution (Figure S2).

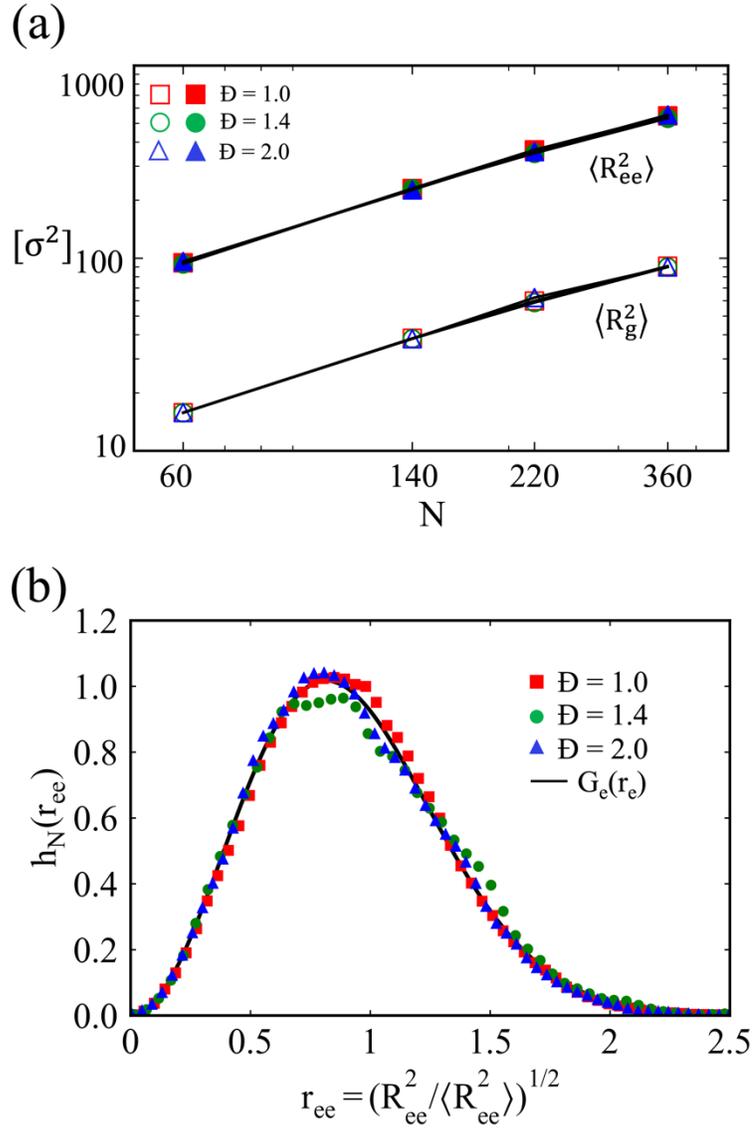

Figure 2. a) Log – log plot of mean square end-to-end distance $\langle R_{ee}^2 \rangle$ and radius of gyration $\langle R_g^2 \rangle$ versus test chain length N in melt of different dispersity, as labelled. The straight line with a slope of 1 indicates a fit of the power law, $\langle R_{ee}^2 \rangle \propto \langle R_g^2 \rangle \propto N^{2\nu}$, $\nu \approx 0.5$. b) Normalized probability distribution of $r_e = \left(R_{ee}^2/\langle R_{ee}^2 \rangle\right)^{1/2}$, $h_N(R_{ee})$ vs. $r_{ee}$ for the same test chain in $M_w = 360$ with Đ = 1.0, 1.4 and 2.0. Solid line shows theoretical predictions $G_e(r_e)$ from Equation 4.

To compute melt structure in a manner consistent with neutron scattering experiments, we compute both the collective S(q) and single – chain static structure factor $S_c(q)$, given by[53]

$$S(q) = \frac{1}{N_{tot}} \langle \sum_{i=1}^{N_{tot}} \sum_{j=1}^{N_{tot}} exp[i\vec{q} \cdot (\vec{r}_i - \vec{r}_j)] \rangle. \quad (5)$$

and

$$S_c(q) = \frac{1}{n_c} \langle \sum_{n=1}^{n_c} \frac{1}{N} \sum_{i,j=1}^{N} exp[i\vec{q} \cdot (\vec{r}_i^{\,n} - \vec{r}_j^{\,n})] \rangle. \quad (6)$$

respectively, where $N_{tot}$ is the total number of monomers (beads) in the melt, $\vec{q}$ is the wave vector, $\langle \cdots \rangle$ is the average over all independent configurations and over all vectors $\vec{q}$ of the same magnitude, $n_c$ is the number of chains, $\vec{r}_i$ and $\vec{r}_j$ are the position vectors of monomer i and j, respectively. S(q) gives a measure of the total scattering from all the monomers in the melt, regardless of whether they are linked along a polymer chain or not. Thus, it represents contributions from intramolecular (corresponding to the average of the standard structure factor, $S_c(q)$ of single polymer chain) and intermolecular interactions. As shown in Figure 3a, we verify the scaling predictions $S(q) \sim q^{-1/\nu}$ at intermediate values, with $\nu = \frac{1}{2}$ for ideal chains, consistent with the results of Hsu and Kremer.[36,53] At the low q regime, for the same $M_w$, S(q) decreases with increasing Đ due to non-uniformity in chain lengths, and increased contribution from short chains. We plot the corresponding Kratky plot in Fig. 3b and compare the results with the Debye function,[54]

$$S_{Debye}(q) = 2\frac{\eta - 1 + exp(-\eta)}{\eta^2} \quad (7)$$

which describes scattering from Gaussian chains. Here, $\eta = q^2 \langle R_{g,Z}^2 \rangle$, $R_{g,Z}$ is the Z-average of the radius of gyration and $N = \langle N^2 \rangle / \langle N \rangle$ in the disperse case.[55] We observe a minimum at some q value, which is an indication of deviation from ideal chains. This minimum becomes less pronounced with increasing Đ, and we attribute the behavior to the increasing degree of non-uniformity in chain length.

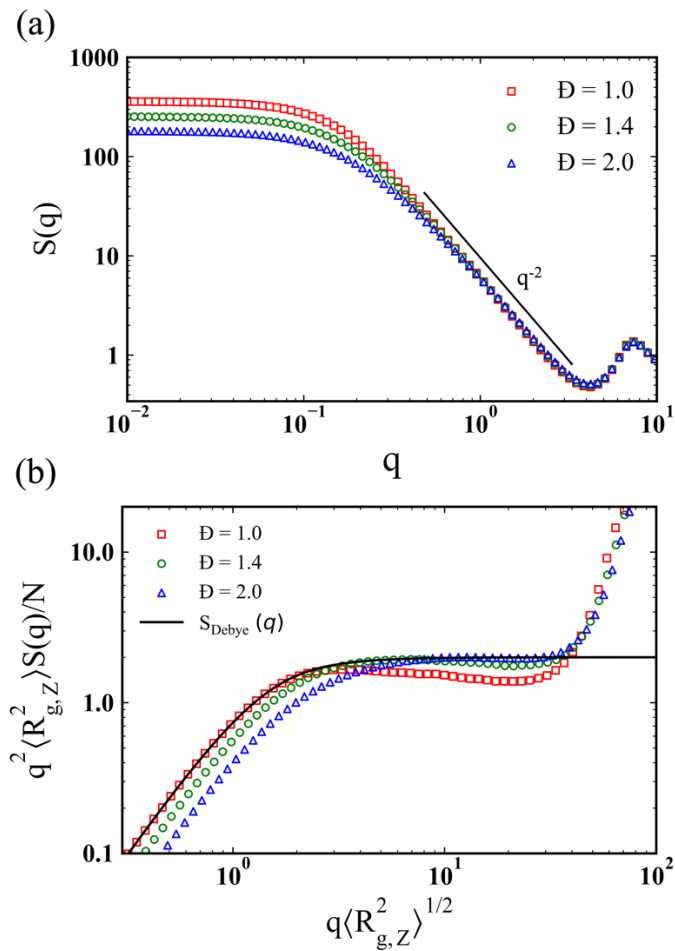

**Figure 3. a) Collective Structure factor S(q) of polymer melts, $M_w$ = 360 with different Đ values as labelled, plotted vs. q on a log – log scale. Scaling of $S(q) \sim q^{-2}$ for Gaussian chains at intermediate q – regime is shown by the solid black line. b) Same data as in (a) but in a Kratky-plot, where $R_{g,z}$ is the Z – average of the radius of gyration and N = $\langle N^2 \rangle / \langle N \rangle$ for the disperse case.**

When we consider scattering from just test chains of size N (Figure 4), we see that $S_c(q) \approx N(1 - q^2 \langle R_g^2 \rangle / 3)$ for small q ($q \ll 2\pi/R_g$, $R_g = \sqrt{\langle R_g^2 \rangle}$), where $\langle R_g^2 \rangle$ is the average mean–squared radius of gyration of the chain). We find that the data collapses on to a single curve for different dispersities, indicating that surrounding chains do not affect the test chain conformation, consistent with previous results of $R_g$ and $R_{ee}$ (Figure 2). At intermediate q values ($2\pi/R_g < q < 2\pi/l_k$), the chains exhibit a power law dependence, behaving as Gaussian coils, with $S_c(q) \sim q^{-1/\nu}$ across all systems (same as previous findings of S(q)). Here, $l_k$ is the Kuhn length, $\nu = 1/2$ and persistence length $l_p = l_k/2$. At large q values ($2\pi/l_k < q < 2\pi/l_b$), the $S_c(q) \sim q^{-1}$ scaling (rigid-rod regime) is absent, indicating fully flexible chains. We also show the corresponding Kratky plot in Figure 4b, and we find that the deviation from Debye scaling is the same for test chains in the melt.

Overall, we see that while computing the average structure of all chains, the deviation from ideality becomes less pronounced with increasing dispersity, however, when we consider test chains of the same size N, the deviation from ideality remains the same irrespective of the heterogeneity of the surrounding chains.

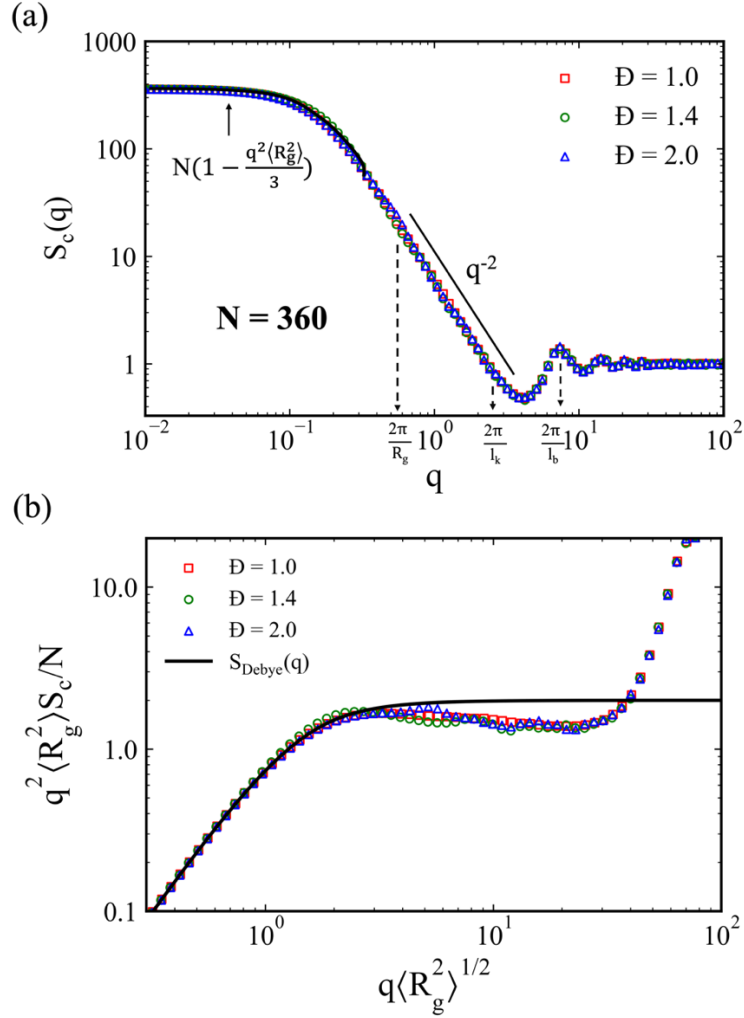

**Figure 4. a)** Single – chain static structure for test chain of N = 360 in different melt, plotted vs. q on a log-log scale. The theoretical predictions of $S(q) \approx N(1 - q^2\langle R_g^2\rangle/3)$ for the Guinier regime at small q values and scaling of $S(q) \sim q^{-2}$ for Gaussian chains is shown by the solid black line. Black dashed lines are used to indicate the crossover points. **(b)** Same data as in (a) but in a Kratky-plot. Solid line indicates Debye function multiplied by $q^2\langle R_g^2\rangle$.

Next, we characterize the dynamic behavior of test chains that are equal to the average molecular weight of the melt (N = $M_w$) through the mean squared displacement of monomers $g_1(t)$, and chain centers of mass $g_3(t)$. This is shown in Figure 5. To minimize the fluctuations caused by monomers at the chain ends, only the MSD of the five middle monomers of each chain were included in the calculations of $g_1(t)$, as done in a previous study.[36] We observe that at early times [for N = 60, t < 10$\tau$ (not shown in plot) and N = 360, t < 3500$\tau$] the monomer mobility remains indistinguishable even at Đ = 2.0 for all cases. We find that the entanglement time $\tau_e$ and length $N_e$ of test chains is independent of dispersity. Overall, at each $M_w$ considered, there is an increase in the test chain mobility with an increase in D. We attribute this increase to an enhancement in the constraint release (CR) mechanism within the tube, which is a topological confinement created by surrounding chains, due to an increase in the number of shorter chains as Đ increases.[56] The CR mechanism is a many-chain effect where the mobility of a specific chain is proportional to chain reptation within a tube. The tube movement itself is Rouse-like and depends on the molecular weight distribution of the melt. Thus, dispersity in chain sizes plays a role in the mobility of test chains, as the obstacles confining the chain in a tube are temporary, existing only as long as the relaxation time of the surrounding chains.[23] In effect, the rapid motion of shorter chains releases the constraints on the longer ones, allowing them to relax more quickly than they would if chain conformation were only dominated by reptation. This is consistent with previous work of Oluseye et. al., albeit for a bidisperse sample and Peter et. al., for narrow molecular weight distribution (low Đ).[4,39] In addition to the test chains, we have also analyzed the dynamics of chains in the monodisperse melts and correlate the different dynamic regimes for the entangled system of $M_w$ = 360 (N = 360) with the scaling theories; this is reported in the Supplementary Information (Figure S4).

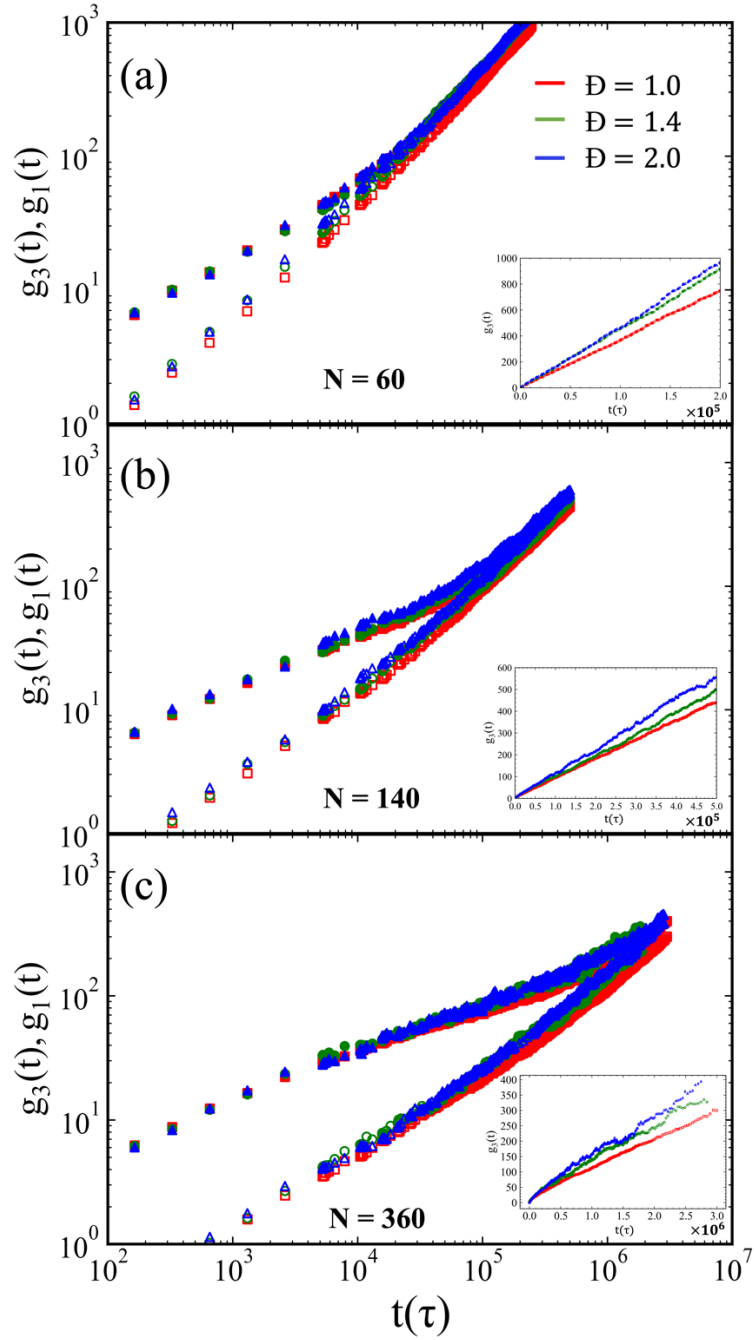

**Figure 5.** Mean squared displacement of the center of mass $g_3(t)$ (open) and monomer $g_1(t)$ (filled) for (a) N = 60 in $M_w$ = 60 (b) N = 140 in $M_w$ = 140 and (c) N = 360 in $M_w$ = 360 for different dispersities, as labelled. Insets show the same data on a linear plot.

To quantify the extent of mobility for test chains at each dispersity, we calculate the self-diffusion coefficient (D) using the Einstein relation $\lim_{t\to\infty} g_3(t) \sim 6Dt$, where D is extracted from the MSD curve, after it has reached a slope of 1 (diffusive regime). The results are shown in Figure 6, and clearly indicate that the test chains with size equivalent to the average molecular weight move significantly faster in a disperse melt than in a uniform melt (i.e., when surrounded by chains of the same size). Interestingly, $\frac{D_{Đ=2.0}}{D_{Đ=1.0}}$, or the ratio of the diffusion coefficients of the test chains in the highest and lowest dispersity melts are 1.4, 3.0, and 1.5 for N = 60, 140, and 360, respectively. This implies that the dynamics of the test chains belonging to the moderate molecular weight system of $M_w$ = 140 has the greatest speedup in mobility due to dispersity, compared to the lowest and highest molecular weight systems. This is because the chains in $M_w$ = 60 system are highly mobile even at low dispersities and increasing dispersity up to 2.0 does not cause a drastic increase in chain mobility. On the other end of the spectrum, for the $M_w$ = 360 system, the increase in dispersity introduces chains that are highly entangled, with the longest chain in the $M_w$ = 360 and Đ = 2.0 system having a size of N = 1600 due to which the short chains can enhance the mobility of the long chains by a small amount. However, in the $M_w$ = 140 systems at higher dispersities, the short chains are able to facilitate constraint release of the longer chains that are only marginally entangled, and thus greatly enhance the mobility of the test chains in the disperse vs. uniform melts.

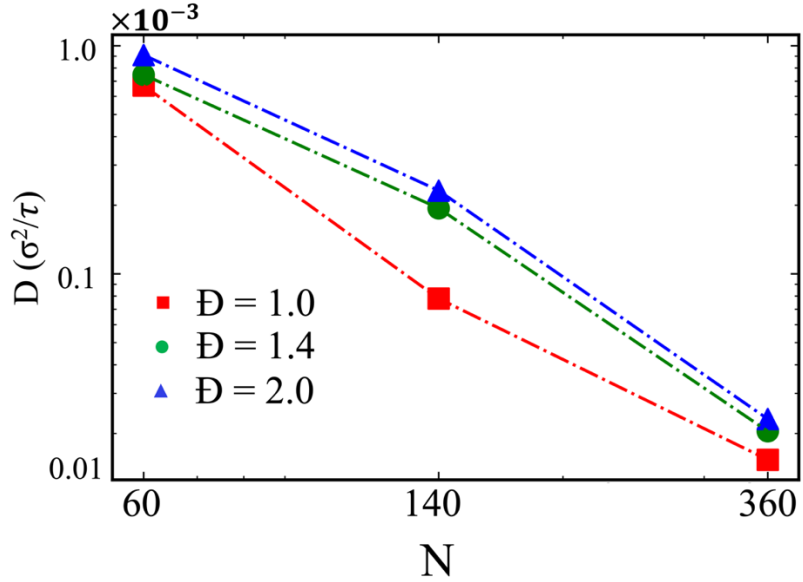

**Figure 6.** Diffusion coefficient as a function of chain length N in different melts, where N = $M_w$, the average molecular weight of the melt. Dash-dotted lines are guides to the eyes.

After having quantified test chains that are equal to the average molecular weight of the melt, we focus on the dynamics of chains higher and lower than the average $M_w$. This allows us to capture the effect of shorter chains on the mobility of chains that are above the weight average molecular weights i.e., $N > M_w$, as well as the influence of longer chains on the mobility of short unentangled chains, $N < M_w$. For the long and short chain analysis, the test chain lengths are N = 500 and 60, respectively, within melts of $M_w$ = 360. The results are shown in Figure 7. From Figure 7a, for the N = 500 test chains, we find that the mobility is independent of Đ at short times, and the monomers have overlapping MSD values. At longer times, the mobility of the test chains increases with dispersity, as the presence of shorter chains accelerates the dynamics of the long test chains. This is not surprising, and consistent with prior literature.[26,39,57] We find that the increase in diffusion constant is linear, as shown in the inset of Figure 7a. The trend is quite different for short test chains, shown in Figure 7b. At long times, the mobility of N=60 chains are the highest in a monodisperse melt, compared to disperse systems. Initially, the MSD of N = 60 chains is lower in the Đ = 2.0 than the Đ = 1.4 system, and at about $3000\tau$, the MSD in the Đ = 2.0 crosses over that of Đ = 1.4 system. At moderate dispersity of Đ = 1.4, the dynamics of the test chains slow down as the influence of long entangled chains outweighs the contribution of the short chains. However, as the ratio of shorter chains is higher at Đ = 2.0 vs. Đ = 1.4 (feature of Schulz-Zimm distribution, Figure 1), it leads to an overall speed up of unentangled chains at long times. We find that the diffusion constant for short test chains is non-monotonic, as shown in the inset of Figure 7b, which underscores how the MWD shape and dispersity can be used to control dynamics of specific chains in the melt.

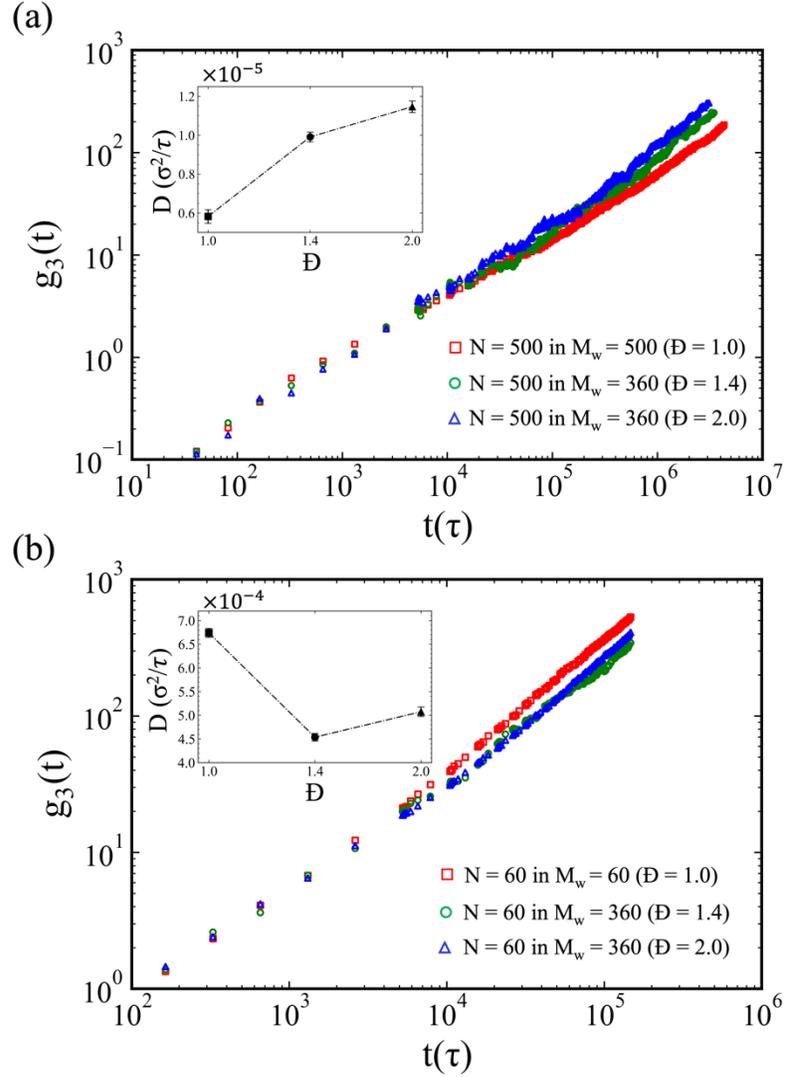

**Figure 7. Mean squared displacement of the center of mass $g_3(t)$ for test chains of length a) N = 500 and b) N = 60 in melts of different dispersities, as labelled. Insets show diffusion coefficient as a function of dispersity. Dash – dotted lines are guides to the eye.**

To quantify chain relaxation similar to experiments, we compute the coherent single chain dynamic structure factor $S_{coh}(q,t)$ defined as[58]

$$S_{coh}(q,t) = \frac{1}{N}\langle\sum_{i=1}^{N}\sum_{j=1}^{N} exp\{i\vec{q} \cdot [\vec{r_i}(t) - \vec{r_j}(0)]\}\rangle. \qquad (8)$$

The average $\langle\cdots\rangle$ denotes an average over all chains, as well as over orientations of the wave vector $\vec{q}$ having the same wavelength. Experimentally, $S_{coh}(q,t)$ can be measured by NSE spectroscopy on a sample containing a small fraction of protonated chains in a deuterated matrix. It has been used to provide microscopic evidence for reptation theory and the two additional relaxation behaviors (CLF and CR) in bidisperse polymer melts.[59] To understand the extent of these relaxation mechanisms in our systems, we report the normalized coherent dynamic structure factor $[S_{coh}(q,t)/S_{coh}(q,0)]$ as a function of time for probe chains that are within ±5 of $M_w$ in melts of different dispersities, as shown in Figure 8. In terms of length and timescales, for $q \ll 2\pi/R_{ee}$, where $R_{ee}$ = end-to-end distance, one can observe the overall diffusion of chain molecules. In the case of N = 140, this corresponds to q << 0.42 and for N = 360, q << 0.27. In general, we find that the decay in $S_{coh}(q,t)$ accelerates with increasing melt dispersity, with lower Đ, corresponding to longer chain disentanglement times, consistent with our previous findings from the MSD analysis. We see that $S_{coh}(q,t)$ at small q values decay slowly for all systems, as this corresponds to chain dynamics at larger length scales. At higher q values (> 0.2), we find that the separation of $S_{coh}(q,t)$ with dispersity is more pronounced in lower molecular weight melts compared to the higher molecular weight melts. This is because the marginally entangled chains of N = 140 are less confined due to the presence of short chains, compared to N = 360 systems which has considerable number of well entangled chains that are not influenced as much by the shorter chains in the melt, leading to greater tube confinement. As a result, the contribution of the well entangled chains leads to a corresponding slowdown in the mobility of the test chains, in the case of N = 360. In comparison with the results of Zamponi et. al.,[29] and Wang and Larson[59] who used NSE spectroscopy and molecular simulations, respectively, to capture the effect of constraint release in binary blends which contain a few long chains in a matrix of short chains, our systems are highly heterogenous, showing a range of dynamic signatures. Overall, the mobility of long test chains in a disperse melt can be highly or moderately enhanced, depending on the width of the distribution and average molecular weight.

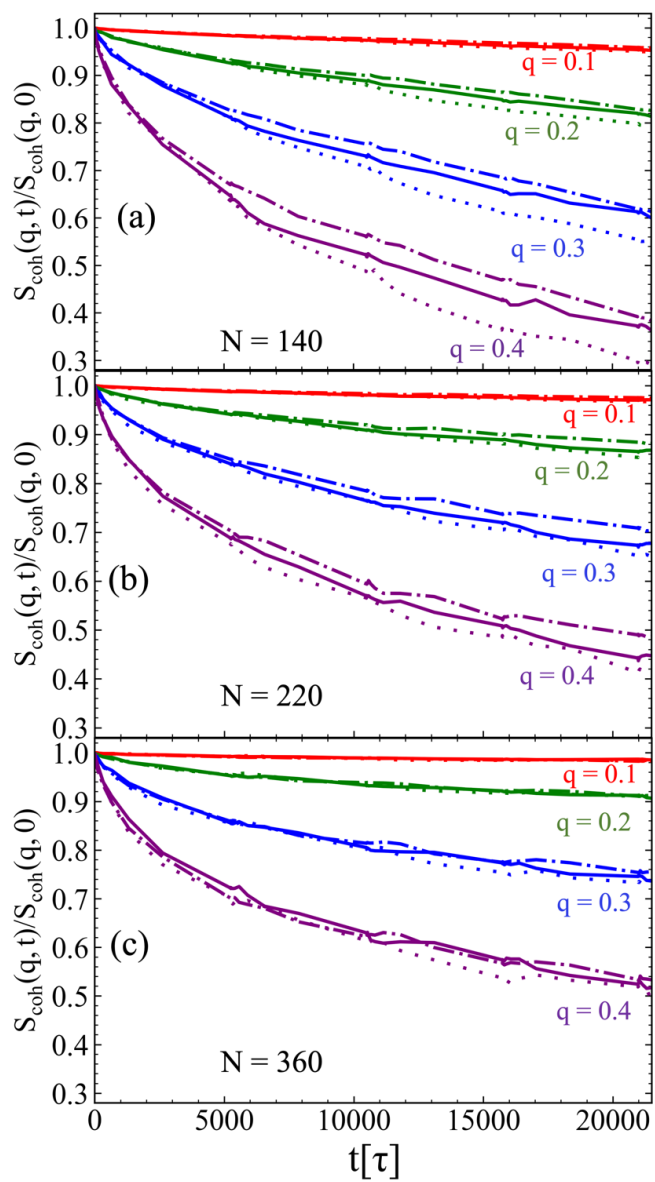

**Figure 8.** Normalized single-chain coherent dynamic structure factors of four q values for test chain lengths of (a) N = 140 in $M_w$ = 140 (b) N = 220 in $M_w$ = 220 (c) N = 360 in $M_w$ = 360. The melt dispersities are Đ = 1.0 (dash-dotted lines), Đ = 1.4 (solid lines) and Đ = 2.0 (dotted lines).

Finally, we calculate the normalized end–to–end vector autocorrelation function $C_{ee}(t)$ to quantify relaxation dynamics of test chains in the melt. We do this by fitting $C_{ee}(t)$ curves with the Kohlrausch - Williams and Watts (KWW) stretched exponential equation[60,61], which is defined as:

$$C_{ee}(t) = e^{\left[-\left(\frac{t}{\tau_{ee}^*}\right)^\beta\right]} \qquad (9)$$

$\beta < 1$ is the stretched exponent parameter, $\tau_{ee}^*$ is the macroscopic time constant also known as KWW characteristic relaxation time. From the autocorrelation, we can obtain the average orientational relaxation time $\tau_{ee}$, as follows:

$$\tau_{ee} = \int_0^\infty e^{\left[-\left(\frac{t}{\tau_{ee}^*}\right)^\beta\right]} dt = \left(\frac{\tau_{ee}^*}{\beta}\right)\Gamma\left(\frac{1}{\beta}\right) \qquad (10)$$

Where $\Gamma(x)$ is a gamma function.

From Figure 9 and as shown in Figure S5 of the supplementary information, we see that the ACF curves decay to zero and the time taken is dependent on system size and Đ. The smallest system $M_w$ = 60, decorrelates fastest and both the $\tau_{ee}^*$ and $\tau_{ee}$ increases with increasing chain size for the monodisperse case (N = $M_w$), in line with prior studies.[28,62] We observe a faster decay of the end – to – end vector as dispersity increases. We ascribe this to an increased contribution of shorter chains (faster relaxation) surrounding the test chains vs. the longer ones (slower relaxation). This leads to a reduction in $\tau_{ee}$ extracted from the time integral equation (inset of the figures).

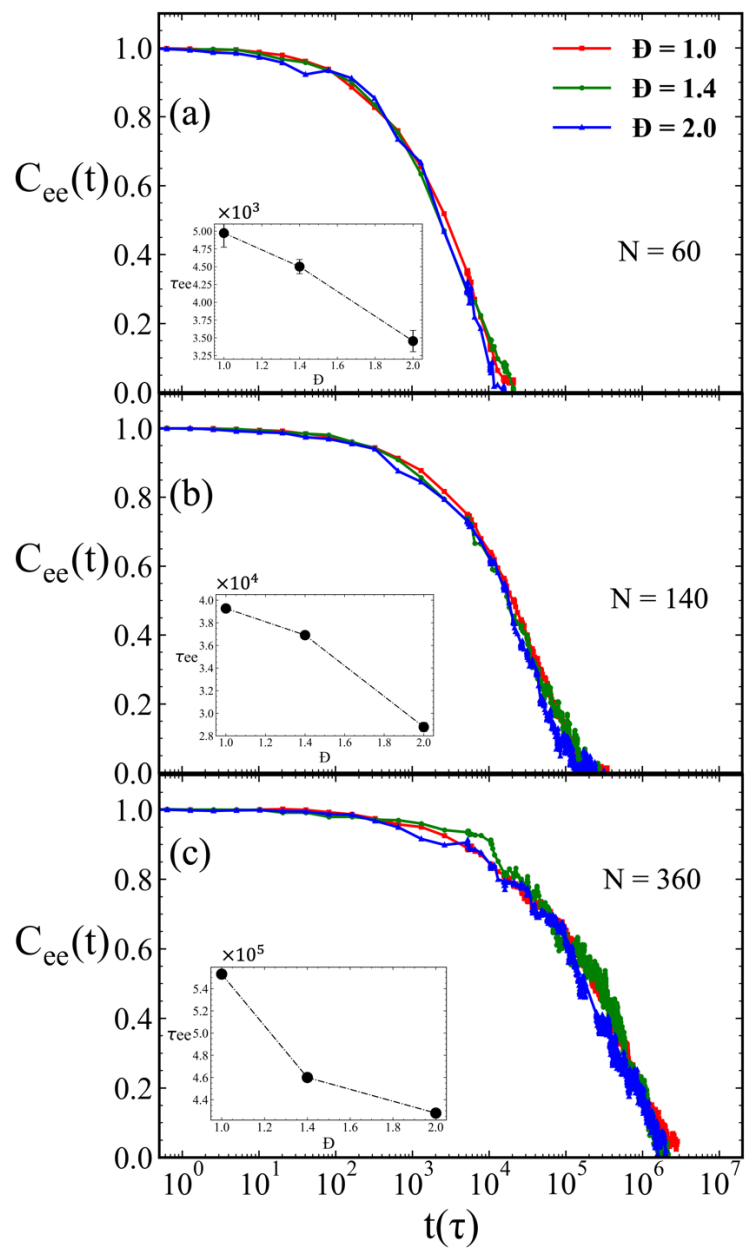

**Figure 9. Normalized autocorrelation function of the chain end-to-end vector $C_{ee}(t)$ of test chain N in melts of different melt dispersities (a) N = 60 in $M_w$ = 60 (b) N = 140 in $M_w$ = 140 and (c) N = 360 in**

**$M_w$ = 360. Inlets show the average chain end-to-end relaxation time $\tau_{ee}$ obtained numerically from time integral of the stretched exponential function. Dash dotted lines are guides to the eye.**

**Conclusions**

In this paper, we have studied the equilibrium conformation and chain dynamics of fully flexible polymer melts that maps Schulz – Zimm molecular weight distribution, by employing coarse – grained molecular dynamics simulation. We show that the global conformational properties of test chains that are equal to the average molecular weight of the melt remain the same irrespective of the heterogeneity of the surroundings. We find that the mobility of test chains that are equal to and greater than the average molecular weight ($N = M_w$ and $N > M_w$) is accelerated as dispersity increases. Interestingly, we see non-monotonic behavior in the diffusion of shorter test chains ($N < M_w$) when both relatively shorter and longer chains are present in a larger melt with varying dispersity. We capture the extent of constraint – release by analyzing coherent dynamic structure factor of the probe chains and observe a faster decay with increasing Ð. This trend becomes less noticeable as the average molecular weight of the system increases, due to a tradeoff in the mobility of chains belonging to distinct dynamical regimes. This interplay of dynamics within the disperse melt, ranging from unentangled to well – entangled chains, provides a critical foundation for capturing microscopic behavior of synthetic polymer melts. Such insights can guide their predictive design for specific applications, such as balancing their viscoelastic and mechanical response, as well as enhancing polymer adhesion.

## Data Availability Statement

Analysis codes, Lammps input script, python script to create initial configuration of the systems can be found here - https://github.com/olajuwonOG/Structure_and_Dynamics_Paper

## Supporting Information (SI)

Supplementary data is available free of charge

## AUTHOR INFORMATION

### Corresponding Author

**Janani Sampath** – Department of Chemical Engineering, University of Florida, Gainesville, Florida 32611, United States

Email: jsampath@ufl.edu

### Authors

**Taofeek Tejuosho** – Department of Chemical Engineering, University of Florida, Gainesville, Florida 32611, United States

**Sohil Kollipara** – Department of Chemistry, University of Florida, Gainesville, Florida 32611, United States.

**Sumant Patankar** - Department of Chemical Engineering, University of Florida, Gainesville, Florida 32611, United States.

### Conflicts of Interest

The authors declare no conflicting interests.

**Acknowledgements**

This project was supported by the department of chemical engineering and Hebert Wertheim College of Engineering at the University of Florida through Professor Sampath's startup fund. The authors acknowledge University of Florida Research Computing for providing computational resources and support that have contributed to the research results reported in this publication. https://rc.ufl.edu/.